\documentclass[11pt,a4paper]{article}

\usepackage[T1]{fontenc}
\usepackage[utf8]{inputenc}
\usepackage{amsmath,amssymb,amsthm}
\usepackage{booktabs}
\usepackage{graphicx}
\usepackage[margin=2.4cm]{geometry}
\usepackage[colorlinks=true,breaklinks=true,linkcolor=blue,citecolor=blue,urlcolor=blue]{hyperref}
\usepackage{microtype}

\newtheorem{lemma}{Lemma}
\newtheorem{proposition}{Proposition}
\newtheorem{corollary}{Corollary}
\theoremstyle{remark}

\newcommand{\tr}{\operatorname{Tr}}
\newcommand{\R}{\mathcal{R}}
\newcommand{\E}{\mathcal{E}}
\newcommand{\Th}{\Theta}
\newcommand{\Msf}{\mathbf{M}}

\newcommand{\maxsd}{\operatorname{maxsd}}

\title{\bfseries Recovery gates as a resource in repetition-code\\[2pt]
process tomography: exhaustive gate-optimal designs\\[2pt]
for three, five and seven qubits}

\author{Long Huang}
\date{%
College of Physics and Electronic Engineering,\\
Sichuan Normal University, Chengdu 610101, China\\[4pt]
\texttt{huangl@sicnu.edu.cn}\\[10pt]
\today}

\begin{document}
\maketitle

\begin{abstract}
Repetition-code tomography compares a code's logical channel with a
reference through a decoder, almost always standard minimum-weight. We ask
how many single-qubit Pauli gates it needs with the measurement basis fixed---the
constraint that makes the count a resource. Exhaustive enumeration at $n=3$, $5$
and $7$, ranks decided in exact arithmetic, gives minima of $0$, $1$ and $5$ gates,
against $6$, $10$ and $14$ for first-order and $6$, $50$ and $308$ for full
decoding. The count depends only on how many sites are corrected, not which, so it
splits into a budget and a placement that decides identifiability; more sites is not
monotonically better: at $n=7$ full decoding has rank $24$ of $27$, and the $n=3$
precision curve is U-shaped under a calibration floor. The three-gate design is
$9.0$ times more precise than the standard six-gate decoder at the median of $40$
working points and better at every one, by $1.8$ to $88$-fold; fixing the placement
in advance gives $6.9$, or $3.1$ for a pre-commitment never worse. Two certified
placements lose no direction either sees and reach $10$\,mrad within twice the shots
wherever identifiable. Measuring the reference everywhere gives a
$16(n+2)$-component map, full rank at $n=3$, $5$ and $7$ with no gate and $44$ to
$93\%$ fewer settings, though at $n=5$ and $7$ its floor stays one and three orders
above target. The optima are not covariant under chain reversal.
\end{abstract}

\section{Introduction and scope}

Reading out the noise of a small register by pushing it through a stabilizer
code and reconstructing the logical channel is an economical alternative to
process tomography of the register itself \cite{chuang,poyatos}: a distance-$n$
repetition code turns
$4^{n}$ measurement settings into a fixed number of syndrome-conditioned
channels, and the two code orientations probe complementary Pauli axes
\cite{companion,huang,lupke}. Repetition-code experiments already characterize
their logical channel in exactly this way, comparing coherent and stochastic
error models at the level of logical fidelity \cite{meer}, and repetitive
syndrome extraction on a chain is itself well established \cite{kelly}. The
logical channel is
produced by three operations that
are not on the same footing. Encoding and propagation are given by the device.
Decoding is a choice---and in practice it is not made, but inherited: the decoder
is taken to be the standard minimum-weight one \cite{nc,dennis,fowler}, because in
error correction that choice maximizes the logical fidelity.

Estimation has a different figure of merit. What matters is not how much of the
error the decoder removes but how much information about the parameters survives
in the channel it produces. These two objectives need not agree, and this note
quantifies how far apart they are. We ask two questions. How many physical
single-qubit Pauli gates does the decoder need? And does correcting more
syndromes buy more information? The answers, for $n=3,5,7$, are $0,1,5$ and no.

\paragraph{The question is only well posed under a constraint.}
The gate count of a decoder can be made to vanish, and it is worth saying exactly
how, because the loophole is easy to fall into. For a Pauli $W_s$ and any Pauli
observable $P$,
\begin{equation}
\tr\!\left(W_s\rho W_s^{\dagger}P\right)
=\tr\!\left(\rho\,W_s^{\dagger}PW_s\right),
\label{eq:frame}
\end{equation}
and $W_s^{\dagger}PW_s$ is again a Pauli. Applying $W_s$ and then measuring $P$
is therefore bit-for-bit the same experiment as measuring $W_s^{\dagger}PW_s$
directly on the uncorrected state. Syndrome-resolved tomography already measures
the syndrome and already scans several bases, so if the basis table may be chosen
per syndrome, the whole decoder is absorbed into that table and costs nothing
\cite{knill}. We therefore impose a \emph{fixed measurement basis}: one
$16$-element Pauli setting table, identical for every syndrome. This is the
regime of a compiled readout basis, where the setting table is calibrated once
and reused across the scan. Under it, $W_s$ cannot be commuted into the basis
choice and must be applied as a physical gate. Every gate count below is a count
in this regime; if the premise is dropped, every count in this note collapses to
zero and the question disappears with it.

\paragraph{What the alternative actually costs.}
The premise deserves more than the one sentence just given, because the
conditional version of it is a real hardware option and, on the numbers below,
the cheaper one is the opposite of what a first reading suggests. Since the
observables are single-qubit Paulis, the conjugation in Eq.~\eqref{eq:frame}
never leaves the site, so a per-syndrome basis table consists of a classical
sign flip---which is free---and a conditional choice between the $X$ and $Z$
bases. The syndrome is known before the final readout in any case, since the
branches are organised by syndrome; what the conditional route needs is a
feed-forward line and one $\pi/2$ pulse selected from the calibrated set. If
the selected pulse is indeed one of the calibrated ones, its error is $p_1$ and
it is paid \emph{once per branch}, not $g$ times. So the two routes cost
\begin{equation}
\text{physical: } g\,p_1,
\qquad
\text{feed-forward: } p_1+p_{\mathrm{ff}},
\qquad
p_{\mathrm{ff}}=\tau_{\mathrm{ff}}/T_2 ,
\label{eq:m1}
\end{equation}
and the comparison is decided by $g$, the average number of Paulis per branch.
It is $0.375$ for the three-gate design at $n=3$---three gates over the eight
code branches---and, more to the point, the maximum over branches is
\emph{one}: the bit code corrects a single syndrome and the phase code two, and
because every corrected syndrome has weight one, no branch applies two Paulis.
Here $g=G/2^{n}$ averages over the $2^{n}$ branches of both codes, and it is the
mean of the two block-level counts $g_{\rm blk}^{(c)}=G^{(c)}/2^{n-1}$ that
Sec.~\ref{sec:precision} needs, since the two blocks are attenuated separately
($0.25$ and $0.5$ here, mean $0.375$). The two readings coincide when the gates
are spread between the blocks and differ by exactly two when they are not: the
optima put every gate in one block, where $g_{\rm blk}=1/16$ at $n=5$ against
$g=1/32$. Every $g$ outside Sec.~\ref{sec:precision} is the design-level one.
That is the whole argument. The physical route costs at most one pulse per
branch, the same as the alternative, and on average less than half of one.
With $\tau_{\mathrm{ff}}=0.5\,\mu$s,
$T_2=100\,\mu$s on the superconducting tables and $T_2=200$\,ms on the ion
table---representative figures, not measured values for any particular device,
and the long ion coherence is why the alternative stays close there---the idle
term is $5\times10^{-3}$ on the former and $2.5\times10^{-6}$ on the latter, so
Eq.~\eqref{eq:m1} gives
$3.8\times10^{-4}$ against $6.0\times10^{-3}$ on sc-cur (a factor $16$),
$1.1\times10^{-4}$ against $5.3\times10^{-3}$ on sc-best (a factor $47$), and
$3.8\times10^{-5}$ against $1.0\times10^{-4}$ on the trapped-ion table (a factor
$2.7$). The physical route wins on all three, and the crossover is explicit: it
wins whenever $g<1+\tau_{\mathrm{ff}}/(p_1T_2)$, a threshold of $1.03$ on the
ion table, $6$ on sc-cur and $17.7$ on sc-best at that latency. At $n=3$ the
standard six-gate decoder sits at $g=6/8=0.75$, so it stays below the ion
threshold as well, by $1.37$. The asymmetry that the argument actually rests on
appears from $n=5$: full decoding needs $50$ and $308$ gates at $n=5$ and $7$,
that is $g=1.56$ and $2.41$ against the $2^{n}$ branches of the two codes, and
it does lose on ions there, while the optima of this paper sit at $g=1/32$ and
$5/128$, that is $0.031$ and $0.039$, thirty-three and twenty-six times below
the same threshold.

What Eq.~\eqref{eq:m1} does not charge the physical route is the latency, and
that omission is worth stating rather than hiding: a Pauli applied once the
syndrome is known waits on the same classical line as a basis switch, so that
idle time is shared and the second term of Eq.~\eqref{eq:m1} is generous to the
alternative. The conclusion survives the stricter accounting. Since $g<1$ for
every design selected here, charge both routes the same $p_{\mathrm{ff}}$ and
the comparison reduces to $g$ against $1$---which every design of this paper
wins, by $1/g$, a factor $2.7$ at the three-gate design. The latency sets the
size of the factors $16$, $47$ and $2.7$, never their sign.
Two things follow, and they are the honest scope of the claim. The gate count
is a resource inside a fixed-basis compilation regime, not a law of nature: a
device that can switch bases per syndrome at low latency turns the count into a
different, and sometimes cheaper, accounting. And the reason the fixed-basis
regime is the interesting one is not that feed-forward is unavailable, but that
the designs which save gates are exactly the ones with $g<1$, where the
physical route is already the cheaper of the two. A design with $g$ well above
one---which full decoding is from $n=5$ onward---would be better compiled the
other way, and that asymmetry is itself an argument for the small designs. At
$n=3$ the two decoders coincide at $g=0.75$, so the distinction is a statement
about $n\ge5$ and not about decoders in general.

\paragraph{Scope.}
We work with the distance-$n$ repetition code on a chain, the small-angle
unitary family of $4n-1$ parameters, and the $48$-component Pauli-transfer
fingerprint \cite{companion}, for $n\le7$; the $16(n+2)$-component extension that
measures the reference channel at every site is treated separately in
Sec.~\ref{sec:level0}. The claims are
of four kinds and are kept apart throughout: exact rank statements at certified
parameter points (Sec.~\ref{sec:rank}), sampled precision over a set of working
points under a stated noise model (Sec.~\ref{sec:precision}), a floor that holds at
every point and requires no choice of placement (Sec.~\ref{sec:floor}), and a cost
model for an actual experiment (Sec.~\ref{sec:cost}). We do not claim a scalability
theorem, a circuit-level threshold, or optimality beyond $n=7$. The experimental
realisation of the three-gate design---syndrome-to-Pauli table, the sixteen
measurement settings, the shot arithmetic on three platform classes, and the
calibration checklist---is the subject of the accompanying supplement
\cite{supp}. The gate counts below do not depend on the archetype---dispersive
$ZZ$ ($4n-1$ parameters) or exchange $XX+YY$ (one more per bond)---since both
reach full rank with the same count at $n=3$ and $n=5$; the comparison between
the two is also in the supplement.

\paragraph{Related manuscripts, and what this one owns.}
Four documents by the same author touch this material, and it is worth saying
which is which so that a reader does not have to acquire the set. The companion
\cite{companion} owns the fingerprint, the reflection ambiguity and its collision
locus, and the $112$-component construction, all at the standard decoder.
Reference \cite{huang} owns the organising view of concatenated three-qubit
codes; a third manuscript (arXiv:2611.11748), on model-family selection and
recovery-gate design in
generator-based noise tomography, fixes a different model family and optimises
the recovery inside it against a narrower objective; where its tables overlap
ours they agree to $0.3\%$, which is a consistency check on a shared fingerprint
and a shared rank engine rather than two independent determinations, and neither
should be cited as evidence for the other. The present note owns the design
problem itself: the exhaustive budget, the position-independence lemma, the
exact rank certificates, the precision ordering, the $\sqrt2$ floor, and the
level-0 second axis of Sec.~\ref{sec:level0}. The
proofs and recorded numbers behind Sec.~\ref{sec:floor} live in a separate
supplementary note \cite{floor}, deposited at Zenodo and cited here by its DOI;
the note and its reproduction package are therefore permanently citable in their
own right, independently of this manuscript's version history. All four are
preprints or under review at the time of writing; none is published, and no
result here relies on one that is.

\section{Model, fingerprint, and the recovery design space}
\label{sec:model}

\subsection{The model and the fingerprint}

With $\alpha_i,\beta_i,\delta_i$ the $X,Y,Z$ angles of site $i$ and
$\gamma_j$ the $Z_jZ_{j+1}$ coupling,
\begin{equation}
K_\theta=\Bigl(\bigotimes_{i=1}^{n}U_i\Bigr)
\exp\Bigl(-i\sum_{j=1}^{n-1}\gamma_jZ_jZ_{j+1}\Bigr),\qquad
U_i=a_i\mathbb{1}-i(\alpha_iX+\beta_iY+\delta_iZ),
\label{eq:model}
\end{equation}
with $a_i=\sqrt{1-\alpha_i^2-\beta_i^2-\delta_i^2}$ and
$\theta\in\Th\subset\mathbb{R}^{4n-1}$ a compact domain on which every $a_i$ is
bounded away from zero. The fingerprint consists of three $4\times4$ Pauli
transfer matrices (PTMs), flattened,
\begin{equation}
\Msf(\theta)=\bigl[\R_{\mathrm{ref}};\R_{\mathrm{bit}};\R_{\mathrm{ph}}\bigr]
\in\mathbb{R}^{48},\qquad
(\R)_{ij}=\tfrac12\tr\!\left[P_i\,\E(P_j)\right],
\label{eq:fp}
\end{equation}
where $\R_{\mathrm{ref}}$ is the channel on site $1$ with the other sites in
$|0\cdots0\rangle$, and $\R_{\mathrm{bit}},\R_{\mathrm{ph}}$ are the logical
channels obtained by encoding, propagating through $K_\theta$, applying a
measurement-free recovery, and decoding. The dimension $48$ is independent of $n$;
only the number of parameters grows.

Adding $\R_0(q_2),\R_0(q_3)$ (level-0 channels on sites $2,3$ with the rest in
$|0\rangle$) gives $\Msf_{80}=[\R_0(q_1);\R_0(q_2);\R_0(q_3);
\R_{\mathrm{bit}};\R_{\mathrm{ph}}]\in\mathbb{R}^{80}$. These need no recovery,
so the parameter count stays $4n-1$; at the reference point rank stays $11$,
$s_{\min}$ rises $2.23\times10^{-2}\to1.22\times10^{-1}$ and $\kappa$ falls
$241\to53$. The three level-0 blocks alone are rank $9$ of those $11$, the two
code blocks supplying the rest;
Sec.~\ref{sec:level0} returns to what this buys, at all three $n$.

\subsection{Recovery as a design, and what it costs}

A recovery design for one code is a pair $(S,w)$: a site set
$S\subseteq\{1,\dots,n\}$ and an order cap $w$, the largest error weight that is
corrected at all. For a syndrome $s$ with minimum-weight representative
$\mathrm{rep}(s)$,
\begin{equation}
W_s=\prod_{i\in \mathrm{rep}(s)\cap S}P_i,\qquad
P_i=\begin{cases}X_i,&\text{bit code}\\ Z_i,&\text{phase code}\end{cases}
\label{eq:Ws}
\end{equation}
and $W_s=\mathbb{1}$ whenever $|\mathrm{rep}(s)|>w$. The number of single-qubit
Pauli gates is the number of syndrome--site pairs,
\begin{equation}
G(S,w)=\sum_{s}\bigl|\mathrm{rep}(s)\cap S\bigr| .
\label{eq:G}
\end{equation}
The family contains the standard decoders: $(S,w)=(\text{all},1)$ is
first-order decoding, $(\text{all},n)$ is full minimum-weight decoding, and
$(\varnothing,\cdot)$ is projection without correction. Two labels that induce
the same per-syndrome site set are the same physical recovery and are counted
once: $(\varnothing,w)$ is one design for every $w$, and $w>\lfloor n/2\rfloor$
coincides with $w=\lfloor n/2\rfloor$ because no representative is heavier than
$\lfloor n/2\rfloor$. With that convention the families hold $8$, $63$ and $382$
designs per code at $n=3,5,7$, and their pairwise products reproduce the
denominators of Table~\ref{tab:rank}---$2002$ pairs at budget five at $n=7$,
$1471$ at budgets up to four. The two codes choose
their designs independently, so a full design is a pair $(S_{\mathrm{bit}},
w_{\mathrm{bit}})\times(S_{\mathrm{ph}},w_{\mathrm{ph}})$ and the total count is
the sum of Eq.~\eqref{eq:G} evaluated twice.

The decisive structural fact is that the count is blind to \emph{which} sites are
chosen.

\begin{lemma}[Position independence of the count]
\label{lem:count}
For the distance-$n$ repetition code and any order cap $w$, the number of
syndromes whose minimum-weight representative has weight exactly $k$ and
contains a given site $i$ equals $\binom{n-1}{k-1}$, independently of $i$.
Consequently
\begin{equation}
G(S,w)=|S|\sum_{k=1}^{\min(w,\lfloor n/2\rfloor)}\binom{n-1}{k-1}
\;\equiv\;|S|\,c_{n,w},
\label{eq:lemma}
\end{equation}
so the count depends on $S$ only through its cardinality.
\end{lemma}

\begin{proof}
A syndrome determines the error up to the code's stabilizer group, and two
distinct errors of weight $\le\lfloor n/2\rfloor$ never share a syndrome in a
code of distance $n$; the minimum-weight representative is therefore unique and
has weight at most $\lfloor n/2\rfloor$. Errors of weight exactly $k$ that
contain $i$ are in bijection with the $(k-1)$-subsets of the remaining $n-1$
sites, hence $\binom{n-1}{k-1}$ of them, and each contributes one gate at site
$i$ by Eq.~\eqref{eq:Ws}. Summing over $k\le\min(w,\lfloor n/2\rfloor)$ and over
$i\in S$ gives Eq.~\eqref{eq:lemma}.
\end{proof}

For full decoding this gives $c_{n}=1,5,22$ at $n=3,5,7$, reproducing the counts
$3,25,154$ per code and $6,50,308$ for both. Table~\ref{tab:rank} uses these
values.

Lemma~\ref{lem:count} is what makes the problem tractable, and it is also its
conceptual content: \emph{cardinality controls cost, position controls
information}. The design problem factorizes into a budget question (how many
sites can I afford?) and a placement question (which ones?), and these can be
optimized in that order. It also explains why gate-saving schemes that only
permute sites---without dropping any---cannot save anything: permuting $S$ leaves
$G$ invariant and only redistributes identifiability.

Two costs sit outside Eq.~\eqref{eq:G} and are design independent: syndrome
extraction, $4(n-1)$ controlled-X gates ($2(n-1)$ checks, two each) over $n-1$
reusable ancillas, and the fixed
$16$-element setting table. Adding them changes no comparison below.

\section{Minimum gate count from exact rank}
\label{sec:rank}

\subsection{Why the rank has to be decided exactly}

Whether a design is admissible is a rank question: the map
$\Msf:\Th\to\mathbb{R}^{48}$ can only be locally inverted where its Jacobian
$J=\partial\Msf/\partial\theta\in\mathbb{R}^{48\times(4n-1)}$ has full column
rank. At $n=7$ this question cannot be answered in double precision, and we
report the failure because it is easy to reproduce. The row norms of $J$ span
$2.1\times10^{21}$, from $2.4\times10^{-21}$ to $5.2$: some fingerprint
components respond
to some parameters at the level of the twenty-first digit. The smallest singular
value of the raw matrix sits at the round-off floor ($\approx10^{-11}$) and the
reported rank then depends on the tolerance rather than on the design \cite{golub}. Row
equilibration repairs the condition number but not the rank: balancing amplifies
the noise along with the weak direction and reads $27$ where the exact value is
$24$. A null-space probe at a rank-deficient design is equally unreliable, since
it inherits the same floor.

We therefore decide the rank in finite fields, and---where the direction of the
implication matters---in $\mathbb{Q}$ itself. The two directions do not cost the
same. A minor that is nonzero mod $p$ is nonzero over $\mathbb{Q}$, so a verdict
of full rank is a proof, and that is the direction the \emph{attainment} half of
every count needs. A minor that \emph{vanishes} mod $p$ proves nothing on its
own: it may be a nonzero integer divisible by $p$. A deficiency read off a
single prime is therefore evidence, not a theorem---and the deficiency is exactly
what the lower bound on $B^{*}$ and the $24/27$ headline rest on.

The modular engine is what makes the sweep possible. With the same even/odd
(split-complex) algebra that produces $J$ over $\mathbb{Q}$ we evaluate every
entry modulo $p$; the bit-code projector is diagonal, the recovery is a
permutation, and the phase-code projector is a Walsh--Hadamard transform, so one
design costs about a second of single-core time at $n=7$. The exact rational
pipeline that rechecks the sweep is the slow one, as it must be: building the
syndrome blocks costs $340$\,s, after which a $27\times27$ rank over $\mathbb{Q}$
costs about $0.3$\,s and the $1471$ pairs take $440$\,s more, against
$0.3$\,s per pair modulo $p$. The certificate alone takes five minutes cold and
under one with the blocks cached. The primary prime is $p=65521$ and every
minimal-count design is rechecked at $p=65519$, so that a submatrix which happens
to vanish at one prime cannot pass for a proof.

Both headline statements are decided over $\mathbb{Q}$ instead, at the certified
in-domain point. That point is rational on the unit sphere and its couplings are
Pythagorean, so every entry of $J$ is rational, and a dual number over
$\mathbb{Q}(i)$ carries the value together with the $27$ derivatives. The rank
of the full-decoding Jacobian comes out at $24$ of $27$, with a certificate that
can be checked without trusting the code:
\begin{itemize}\itemsep1pt
\item a $24\times24$ minor whose determinant is a nonzero integer of $2519$
  digits, so the rank is at least $24$;
\item three primitive integer kernel vectors, entries below $2^{4700}$, each
  satisfying $Jv=0$ entry by entry in integer arithmetic, so the rank is at most
  $24$.
\end{itemize}
\noindent Equality, not an estimate, and a second in-domain point with unrelated
denominators reproduces it: rotations with denominators $7814$, $62509$ and
$62517$ and couplings $84/1763$, $96/2303$, $104/2703$, $116/3363$, $124/3843$,
$132/4355$, sharing no factor with the first point, again give rank $24$. The
kernel vectors are dense---$25$ of $27$ entries
are nonzero---so the deficiency is not a symmetry of the model that could have
been spotted by hand. The same construction, with the syndrome blocks cached
once, decides the sweep as well: all $1471$ design pairs at $B\le4$ have rank at
most $26$ over $\mathbb{Q}$ (the maximum is $14$, $19$, $22$, $23$ and $26$ at
$B=0,\dots,4$), so $B^{*}=5$ is a lower bound in exact arithmetic and not only
modulo $p$---and the exact rank equals the modular one for every one of the
$1471$ pairs. The sweep was then repeated at the second certified point, the one
with the unrelated denominators named above, and it returns the same maximum at
every budget---$14$, $19$, $22$, $23$ and $26$ for $B=0,\dots,4$, with rank $24$
at full decoding---so the lower bound on $B^{*}$ does not depend on which of the
two points is chosen. That is the recheck the rank statements elsewhere in this
note get; the deficiency is a property of the design family at these points, not
of one arithmetic accident, although it remains a statement about points in
$\Th$ rather than about all of $\Th$ (Sec.~\ref{sec:limit}). The
two pipelines differ in where they compute: one evaluates in $\mathbb{Q}(i)$ and
reduces mod $p$ to eliminate in $\mathbb{F}_p$, the other stays in $\mathbb{Q}$
throughout. They agree row by row at $n=7$ as well ($0/48$ mismatches), a
stronger cross-check than the one available at $n=3$ and $5$.

\subsection{Results}

The sweep is exhaustive by budget: for each gate count $B=0,1,2,\dots$ every
design pair with $G_{\mathrm{bit}}+G_{\mathrm{ph}}=B$ is examined, and the
smallest $B$ at which a full-rank pair exists is the minimum $B^{*}$.

\begin{table}[t]
\centering
\caption{Gate counts and exact ranks for the three-component fingerprint
($48$ components); the $16(n+2)$-component extension of
Sec.~\ref{sec:level0} has $B^{*}=0$ at all three $n$ at the certified point. $B^{*}$ is the smallest
total gate count
at which \emph{any} design pair is full rank. The fourth column reads
$F/A$: $F$ pairs of rank $4n-1$ among the $A$ whose combined gate count is
\emph{exactly} that budget---for $n=7$ a further $1471$ have fewer gates, so $A$
is not cumulative. The fifth repeats the test at the second prime, $p=65519$,
and reads $F'/F$. ``First order'' is $(\text{all},1)$, ``full'' is
$(\text{all},n)$. The last column is the rank at full decoding, which is where
the count stops being a proxy for information; it is decided in exact rational
arithmetic, not modulo a prime.}
\label{tab:rank}
\begin{tabular}{lccccccc}
\toprule
$n$ & params & $B^{*}$ & full rank / pairs & also at $p=65519$
& first order & full & rank at full\\
\midrule
$3$ & $11$ & $\mathbf{0}$ & $1/1$ & $1/1$ & $6$ & $6$ & $11/11$\\
$5$ & $19$ & $\mathbf{1}$ & $5/10$ & $5/5$ & $10$ & $50$ & $19/19$\\
$7$ & $27$ & $\mathbf{5}$ & $945/2002$ & $945/945$ & $14$ & $308$ & $\mathbf{24/27}$\\
\bottomrule
\end{tabular}
\end{table}

Table~\ref{tab:rank} is the main result of this section. Three points deserve
emphasis.

\emph{The minimum is small and it is proved, not sampled.} At $n=7$ there is no
full-rank design at $0,1,2,3$ or $4$ gates---every pair in each of those budgets
was examined---so $5$ gates is a lower bound as well as an attainment. The
saving in gate count over full decoding is $98.4\%$, and it is not paid for in
rank: the five-gate design is full rank, whereas full decoding is not.

\emph{The two codes are not interchangeable.} At $n=5$, correcting a single
syndrome on the phase side leaves the map full rank for any of the five sites;
correcting a single syndrome on the bit side leaves rank $14$ or $15$ out of
$19$ for all five. The same gate, spent on the same site, buys nothing on one
side and everything on the other. This is why designs that force both codes to
share a site set overpay: at $n=7$ such designs need three shared sites ($6$
gates) to reach $27/27$, since no pair with two shared sites, i.e.\ four gates,
is full rank at all. Decoupled, five gates suffice, and only split $3+2$ or
$2+3$: of the $945$ full-rank pairs at that budget, $420$ correct three bit
sites and two phase sites, $525$ the other way. One of the $420$ is bit
$\{3,5,6\}$ with phase $\{1,6\}$, the design whose condition number is quoted in
Sec.~\ref{sec:precision}. It is a representative, not \emph{the} design: what
Lemma~\ref{lem:count} fixes is the count, not the identity, and bit $\{1,2,3\}$
with phase $\{5,6\}$ costs five gates too but has rank $26$ of $27$. Decoupling the two designs is not a cosmetic degree of freedom; it is
where the saving comes from.

\emph{More decoding can mean less information.} At $n=7$, full minimum-weight
decoding has rank $24$ of $27$---exactly, over $\mathbb{Q}$, by the certificate
of Sec.~\ref{sec:rank}, not merely modulo a prime. At this point the three blocks
have ranks $3+9+12=24$ and their row spaces intersect trivially in pairs, so no
two blocks share a direction that could compensate for a direction the third one
misses, and the total is pinned to the sum. The five-gate design reaches $27$,
which is only possible if its blocks do share directions; it does not partition
the space the way full decoding does. Correcting more syndromes pushes the
fingerprint towards a decoder-determined subspace and destroys the redundancy
that made the parameters separable. Rank is not a monotone function of effort,
and this is the mechanism behind the non-monotonicity that
Sec.~\ref{sec:precision} finds in precision.

The number is attached to the point, and only downwards: rank is lower
semicontinuous, so a degenerate point can lose more than a generic one. One with
several vanishing rotation components---outside $\Theta$, but the kind of point a
scan aligned with the axes passes through---gives rank $23$. On the domain the
deficiency is three, and the mechanism above does not change with the count.

\section{Precision under a fixed-basis noise model}
\label{sec:precision}

Rank says a design is admissible; it says nothing about how well the parameters
are determined. Two designs can both be full rank and differ by two orders of
magnitude in the error they deliver, because they weight the weak directions
differently.

\subsection{Noise model}

Each fingerprint component is a PTM entry with $|R_{ij}|\le1$, measured from
$N$ repetitions of a branch, so its shot error falls as $N^{-1/2}$
\cite{giovannetti}. A branch applies $g_{\rm blk}$ recovery gates on average, where
$g_{\rm blk}=G^{(c)}/2^{n-1}$ with $G^{(c)}$ the count of Eq.~\eqref{eq:G} for
\emph{one} code---the two blocks carry their own $g_{\rm blk}$, $0.25$ and $0.5$
at the quoted design---and the reference branch, which applies none, is not in the
average. This is twice the design-level $g$ of Eq.~\eqref{eq:m1} at the optima,
which place every gate in one block; the two agree only for a decoder that spreads
its gates evenly, so the substitution matters wherever a floor is quoted. Random gate,
readout and preparation errors only dim the signal,
\begin{equation}
\eta(g_{\rm blk})=(1-2p_1)^{g_{\rm blk}}\,(1-2p_2)^{4(n-1)}\,(1-2p_{\mathrm{ro}})(1-p_{\mathrm{prep}}),
\label{eq:eta}
\end{equation}
in which the second factor is the $4(n-1)$ controlled-X gates of the extraction:
$2(n-1)$ checks, each a two-controlled-X parity measurement, one set for the bit
code and one for the phase code---the same $8$ at $n=3$ that
Sec.~\ref{sec:cost} charges in the circuit time. It is the one factor here that
carries no design index, and that is worth a sentence: the checks are the same
for every branch of every design, so $p_2$ multiplies all three blocks by one
constant and cancels from every ratio in this note. It does not cancel from the
absolute budgets, and there it is not negligible---$(1-2p_2)^{8}=0.938$ on
sc-best, $0.984$ on ion, $0.851$ on sc-cur. A budget scales as $1/\eta^2$, and
the floor moves with $\eta$ besides, so restoring $p_2$ raises the sc-best
budget by $19\%$---$1.27\times10^{8}$ becomes $1.51\times10^{8}$ and $707$\,s
becomes $840$\,s---and the ion budget by $3\%$, where the floor sits far enough
below the target for the shot term alone to set the scaling.
while calibration residuals---drift, crosstalk, model error---enter as a variance
floor that more shots cannot remove,
\begin{equation}
\sigma^2=\frac{1}{N}+\sigma_{\mathrm{res}}^2 .
\label{eq:sigma}
\end{equation}
The two are kept apart deliberately: $\eta$ is a hardware property that shifts
the whole fingerprint, $\sigma_{\mathrm{res}}$ is a systematic effect that limits
the noise floor of the estimate. A floor of this kind is not removed by
oversampling---that is the lesson of self-consistent process tomography and of
gate-set tomography, where the error of the gate used to interrogate the system
is itself unknown \cite{merkel,blumekohout}---and it is why we quote it
separately from the shot term throughout. The parameters are then extracted by generalized
least squares, $\mathrm{Cov}=(\tilde J^{T}\Sigma^{-1}\tilde J)^{-1}$ with
$\tilde J=\mathrm{diag}(\eta)J$ and $\Sigma=\mathrm{diag}(\sigma^2)$ \cite{kay}, and we
quote the rms over \emph{all} $4n-1$ parameters,
$\sqrt{\tr\mathrm{Cov}/(4n-1)}$, together with a balance figure
$\max_i d_i/\operatorname{med}_i d_i$ ($d_i$ the marginal standard deviation).
The rms criterion is the one that matches the purpose: a single blind direction
means the tomography is not finished, however good the other parameters look.
The balance figure says whether the design spreads its precision evenly or
concentrates it.

Platform parameters are representative, not device-specific: the single-qubit
error $p_1$, the controlled-X error $p_2$, the readout and preparation errors,
and the residual floor $\sigma_{\mathrm{res}}$ take the values
$(10^{-3}$, $10^{-2}$, $10^{-2}$, $10^{-3}$, $10^{-3})$ for a current
superconducting device (sc-cur), $(3\times10^{-4}$, $4\times10^{-3}$,
$5\times10^{-3}$, $3\times10^{-4}$, $3\times10^{-4})$ for a best-case one
(sc-best), and $(10^{-4}$, $10^{-3}$, $10^{-3}$, $10^{-4}$, $10^{-4})$ for a
trapped-ion device (ion), with cycle times $5\,\mu$s and $200\,\mu$s
respectively. The cycle time is not the whole of a shot: Sec.~\ref{sec:cost}
costs a shot as $t_{\mathrm{cycle}}$ plus the circuit time of the branch.

\paragraph{What Eq.~\eqref{eq:eta} assumes, and what it costs to assume it.}
Two idealisations sit inside Eq.~\eqref{eq:eta}, and both are stated here rather
than left to be inferred. The first is that the readout factor is a visibility
of the measured Pauli, applied once per component. The second is that the
syndrome is resolved exactly, so that a shot belongs to the branch it is filed
under. The second is the one that costs something, and it does not behave like a
dimming: if an ancilla is misread, a shot of syndrome $s$ is multiplied by the
recovery $W_{s'}$ of a neighbouring label $s'$ rather than by $W_s$, so the
block receives $V^{\dagger}W_{s'}P_sK_\theta V$ in place of
$V^{\dagger}W_sP_sK_\theta V$. Summing over the true syndrome, the displacement
is at most
\begin{equation}
\|\delta\R_{c}\|\le 2\,(n-1)\,p_{\mathrm{ro}}\,\|\R_{c}\| ,
\label{eq:crosstalk}
\end{equation}
linear in $n-1$ because each of the ancillas is an independent way to mislabel
the shot. Evaluated on the quoted three-gate design the coefficient is not a
small correction: $(n-1)p_{\mathrm{ro}}$ is $2.0\times10^{-2}$ on sc-cur,
$1.0\times10^{-2}$ on sc-best and $2.0\times10^{-3}$ on ion, and the displacement
computed from that leakage sum---where a misread syndrome swaps one recovery for
a neighbour's, so the full block difference is replaced by an average over the
possible neighbours---is
$1.0\times10^{-2}$, $5.0\times10^{-3}$ and $1.0\times10^{-3}$: one order of
magnitude \emph{above} the calibration floor $\sigma_{\mathrm{res}}$ of
$10^{-3}$, $3\times10^{-4}$ and $10^{-4}$ on the same three tables. Readout
error, not calibration drift, is therefore the binding systematic of the
protocol as modelled, and the honest reading of Eq.~\eqref{eq:eta} is that it
describes a device on which readout has been characterised and mitigated---a
confusion-matrix unfolding, standard in this literature \cite{nachman}---not a
device on which
it has been ignored. We quantify it rather than absorb it because a $10$\,mrad
target is only meaningful once it is clear which systematic sets the floor, and
this one is design independent: like $p_2$ it multiplies every design by the
same constant, so it moves no comparison in Secs.~\ref{sec:precision}
and~\ref{sec:cost} but it does bound what they can promise.

\subsection{Three qubits: a U-shaped curve}

\begin{table}[t]
\centering
\caption{$n=3$, precision against gate count. Medians and rankings are over $40$
working points (one certified exact point plus random points with
$\lambda\in[0.005,0.05]$); the last two columns are a single canonical point at
$N=10^{7}$ shots per component per branch, i.e.\ $1.44\times10^{9}$ in total.
Each row reports, at every working point, the best \emph{placement} available at
that budget; the budget itself is the row label. The six-gate budget needs no
such qualification, because it contains exactly one design pair---every site of
both codes, $c_{3}=1$ per site---so that row \emph{is} the standard decoder
$(\text{all},1)$ and has no placement freedom to exercise; the asymmetry the
reader should keep in mind is therefore not between rows but between the
three-gate rows, where the placement matters, and the six-gate row, where it
cannot. ``Wins'' counts the working points at which the best design of that
budget attains the smallest rms of the seven. The count is sensitive to
near-degeneracy---where two gates wins it is ahead of three by less than
$1.5\times10^{-3}$ relative, where four gates wins by less than
$3\times10^{-2}$---so single wins are not a stable statistic, and the median and
the mean rank are the columns to read. The canonical point is a certified exact
point whose six-gate Jacobian is conditioned $1.6$ times worse than the median
of the sampled set, so the two absolute columns are conservative rather than
typical.}
\label{tab:n3}
\begin{tabular}{cccccc}
\toprule
gates & median rms (rad) & wins & mean rank & canonical rms & balance\\
\midrule
$0$ & $5.43\times10^{-1}$ & $0/40$ & $6.80$ & $6.36$ & $8.51$\\
$1$ & $1.11\times10^{-1}$ & $0/40$ & $4.45$ & $5.57\times10^{-2}$ & $6.15$\\
$2$ & $3.58\times10^{-2}$ & $9/40$ & $2.50$ & $2.62\times10^{-2}$ & $4.72$\\
$\mathbf{3}$ & $\mathbf{3.00\times10^{-2}}$ & $\mathbf{17/40}$ & $\mathbf{1.60}$
& $\mathbf{1.68\times10^{-2}}$ & $\mathbf{1.68}$\\
$4$ & $3.20\times10^{-2}$ & $14/40$ & $1.95$ & $1.68\times10^{-2}$ & $1.68$\\
$5$ & $1.16\times10^{-1}$ & $0/40$ & $4.55$ & $9.95\times10^{-2}$ & $28.66$\\
$6$ & $2.84\times10^{-1}$ & $0/40$ & $6.15$ & $3.75\times10^{-1}$ & $2.71$\\
\bottomrule
\end{tabular}
\end{table}

\begin{figure}[t]
\centering
\includegraphics[width=\linewidth]{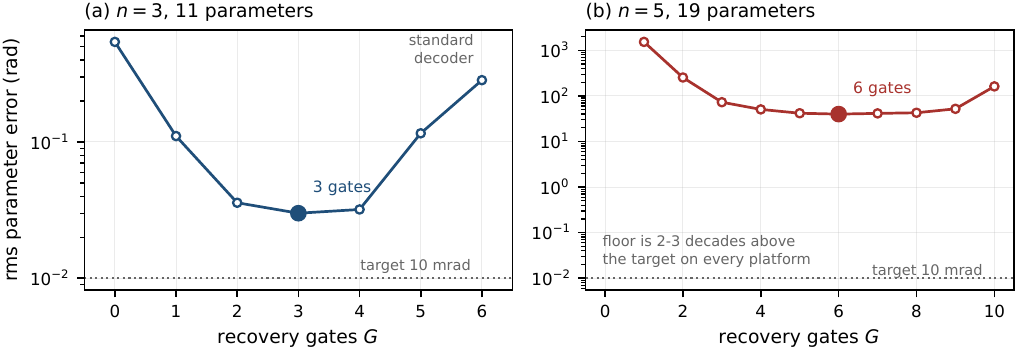}
\caption{Precision against recovery-gate count, medians over the working points.
(a) $n=3$: the curve is U-shaped with a minimum at three gates; the standard
six-gate decoder is $9.0$ times worse at the median, and worse at every point
($1.8$ to $88$ times) once the placement is chosen per point. (b) $n=5$: a broad plateau around six
gates, but the whole curve, floor included, lies two to three decades above the
$10$\,mrad target on every platform. Filled markers are the quoted designs.}
\label{fig:curve}
\end{figure}

Table~\ref{tab:n3} and Fig.~\ref{fig:curve} show the result. The curve is
U-shaped, not monotone. Zero gates keeps the map full rank but is useless for
estimation: its rms is $18$ times worse than the three-gate design, and at the
canonical point $379$ times worse. Two gates is a genuine competitor ($9/40$
wins) but its balance is $4.72$: it concentrates the precision on a subset of
parameters. Three gates wins $17/40$, places in the top three at all $40$, and
has the best mean rank and the flattest balance ($1.68$); those $17$ wins are
shared among placements, with bit $\{3\}$ and phase $\{1,2\}$ accounting for
$11$ of them. Four gates is numerically indistinguishable from three at the
canonical point ($1.68\times10^{-2}$ and $1.68$) and we quote the cheaper design.
Five gates degrades sharply---balance $28.66$, i.e.\ two parameters nearly
collapse---and the standard six-gate decoder, the design a reader would have used
by default, is $9.0$ times worse than three gates at the median of the per-point
ratios ($9.5$ between the two medians) and worse at all $40$ points, by $1.8$ at
the least favourable and $88$ at the most. The ratio is right-skewed (mean $12$,
fifth percentile $2.5$), so the median is the conservative summary of the gain,
not the flattering one.

The $9.0$ is the gain of choosing the placement per point, not of a fixed recipe.
Pre-committing to bit $\{3\}$, phase $\{1,2\}$ retains $6.9$ at the median but is
worse than full decoding at two of the $40$ points, by up to $2.8$; pre-committing
to bit $\{1,2\}$, phase $\{3\}$ retains $3.1$ and is never worse. The experimenter
who must build the decoder before the parameters are known is trading a factor $2$
of median gain for the elimination of the tail, and neither choice delivers the
$9.0$ that per-point optimisation reports.

\paragraph{What the ordering is a statement about.}
The factor comes from the geometry of the two Jacobians and hardly at all from
the noise table, and it is worth separating the two. The floor
$\sigma_{\mathrm{res}}$ is common to all $48$ components, so changing it
multiplies both covariances by the same constant: scaling it by $0.3$ and by $3$,
removing it altogether, scaling the reference block's floor alone by the same
factors, and swapping all four platform tables all leave the median ratio at
$6.9$ (quoted placement) and $9.0$ (placement chosen per point) to one decimal.
What would move it is a change of \emph{shape}. If each component were noisy in
proportion to its own scale rather than with the same absolute $\sigma$, the
ratio falls to $1.0$ at the median and the three-gate design wins at $16$ of the
$40$ points. The absolute model is the physical one here---a PTM entry is a
frequency whose shot error does not shrink with the entry---so the ordering holds
under the model that describes the experiment, but the dependence deserves to be
stated: the $9.0$ is a statement about absolute parameter error, and a
relative-error figure of merit would not support it.

\paragraph{Why the binomial model is not the intermediate case.}
An absolute error $1/N$ is an upper bound, not the variance: a component with
true value $R_{ij}$ estimated from $N$ shots has
$\sigma_{ij}^{2}=(1-R_{ij}^{2})/N$, which vanishes as $|R_{ij}|\to1$. It is
therefore natural to expect the truth to lie between the absolute model and the
proportional one. We ran that model,
$\sigma_{ij}^{2}=(1-(\eta R_{ij})^{2})/N+\sigma_{\mathrm{res}}^{2}$, over the
same $40$ points and the same designs, and it lands on top of the absolute one,
not between the two:
\begin{center}
\begin{tabular}{lccc}
\toprule
model & median $r$ (quoted) & median $r$ (per point) & wins\\
\midrule
absolute, $1/N$ & $6.93$ & $9.01$ & $38/40$, $40/40$\\
binomial, $(1-R^{2})/N$ & $7.00$ & $9.08$ & $38/40$, $40/40$\\
proportional & $1.0$ & $1.1$ & $16/40$, $37/40$\\
\bottomrule
\end{tabular}
\end{center}
\noindent and it does not move the minimum either: the median rms at three
gates is $3.01\times10^{-2}$ under the absolute model and $2.99\times10^{-2}$
under the binomial one, against $3.19\times10^{-2}$ at four gates. The reason is
a property of the fingerprint rather than of the estimator. Of the $48$
components, $12$ have $|R_{ij}|>0.99$---they are the identity entries and the
near-identity diagonal of the reference block---and those carry almost no
parameter information; the median of $1-R_{ij}^{2}$ over all $48$ is $1.000$ and
the mean is $0.75$. The components that determine the covariance are thus
already at the noise ceiling $1/N$, and the components that the binomial model
quiets are the ones that were contributing nothing. What would move the ratio
is a model in which the \emph{informative} components are quieter than the
uninformative ones, which is the proportional model, and it moves the ratio to
$1$. So the sensitivity is real but it is a sensitivity to a sign, not to a
degree: any model that quietens the strong components and leaves the weak ones
alone falls to $1$, and any model that does not leaves the ordering at $9$.
The binomial scan is therefore reported as the referee-relevant check rather
than as a correction: it changes no quoted number by more than $1\%$.

The mechanism is the one identified in Sec.~\ref{sec:rank}. Adding recovery sites
does not monotonically add information: it first fills in directions the
fingerprint is missing, and then starts to project the fingerprint onto a
decoder-selected subspace, flattening the weak direction. The optimum sits where
those two effects balance, and it is not at the maximum of the second one.

\subsection{How much of this belongs to the working point}
\label{sec:point}

The reference point at which the canonical columns of Table~\ref{tab:n3} and all
of Sec.~\ref{sec:cost} are evaluated is a certified exact point, chosen so that
the numbers can be reproduced without sampling. It is not a favourable one: its
six-gate Jacobian has condition number $6.4\times10^{3}$ against a median of
$4.0\times10^{3}$ over the $40$ sampled points, so those columns are conservative
by a factor $1.6$ in conditioning. The companion paper's working point sits at
the other end---$|\theta|\lesssim0.1$\,rad, condition number
$1.3\times10^{2}$, a factor $4.6$ below the best of the $40$ sampled points,
whose values run from $5.9\times10^{2}$ to $3.6\times10^{5}$.

\begin{table}[t]
\centering
\caption{The same three designs at two working points. $A$ is bit $\{3\}$, phase
$\{1,2\}$; $B$ is bit $\{1,2\}$, phase $\{3\}$; ``full'' is
$(\text{all},\text{all})$. Columns 3--5 are the shot-independent floors in rad;
the last column is the total shot budget and wall clock for a $10$\,mrad rms on
sc-cur, with ``---'' when the floor is already above the target. $\kappa$ is the
condition number of the full-decoding Jacobian at that point.}
\label{tab:points}
\begin{tabular}{llccccc}
\toprule
point & design & sc-cur & sc-best & ion & $N_{\mathrm{tot}}$ at $10$\,mrad\\
\midrule
reference      & $A$ & $1.66\times10^{-2}$ & $4.91\times10^{-3}$ & $1.62\times10^{-3}$ & ---\\
$\kappa=6.4\times10^{3}$ & $B$ & $4.08\times10^{-2}$ & $1.21\times10^{-2}$ & $4.00\times10^{-3}$ & ---\\
               & full & $3.71\times10^{-1}$ & $1.10\times10^{-1}$ & $3.63\times10^{-2}$ & ---\\
\midrule
companion      & $A$ & $6.36\times10^{-3}$ & $1.89\times10^{-3}$ & $6.24\times10^{-4}$ & $2.5\times10^{7}$ (138\,s)\\
$\kappa=1.3\times10^{2}$ & $B$ & $4.93\times10^{-3}$ & $1.46\times10^{-3}$ & $4.83\times10^{-4}$ & $1.2\times10^{7}$ (65\,s)\\
               & full & $1.11\times10^{-2}$ & $3.30\times10^{-3}$ & $1.09\times10^{-3}$ & ---\\
\bottomrule
\end{tabular}
\end{table}

Table~\ref{tab:points} puts the two side by side. Three things move. The floors
move, by a factor $33$ at full decoding and $3.4$ at three gates. The better of
the two three-gate placements flips: $A$ leads by $2.5$ at the reference point,
$B$ by $1.3$ at the companion point, with identical hardware and only the look-up
table changed. And the margin of three gates over six narrows from $22$ to $2.3$.

One thing does not move: the ordering. Three gates is the minimum of the floor at
both points; four gates ties it at the reference point (Sec.~\ref{sec:cost}) and
is within $20\%$ at the other; and full decoding is the worst design in the
scanned budget range at both.

The reason is the condition number, and its reach is worth stating. Over the
sampled domain $\kappa(J_{\mathrm{full}})$ spans $5.9\times10^{2}$ to
$3.6\times10^{5}$, a factor $600$---more than any change of gate count within the
budget buys, since the whole three-to-six comparison is a factor $22$ at its most
favourable. A floor quoted for a design is therefore a statement about a
neighbourhood of one working point, and the statement that transfers is the
comparison at fixed point.

\subsection{Five and seven qubits}

At $n=5$ the optimal budget is a plateau: $5$--$8$ gates all give a median rms of
$40$--$43$\,rad, with six gates winning $4/10$ points and having the best mean
rank ($2.40$). The minimum \emph{gate count} of one (Table~\ref{tab:rank}) is a
different question, and answering it with the rms criterion is instructive: the
one-gate design has a median rms of $1.5\times10^{3}$ rad and, at the canonical
point, $3177$ rad against $19.2$ for six gates---a factor $165$. The one-gate
design is the minimum for \emph{identifiability}, and it is unusable for
\emph{estimation} because it abandons the weak parameters. The two criteria
answer different questions and we report both rather than choosing one.

More important, the entire $n=5$ curve is out of reach. Even the best five-gate
design---bit $\{4,5\}$, phase $\{1,4,5\}$---has a shot-independent floor of
$18.9$ rad (sc-cur), $5.60$ rad
(sc-best) and $1.85$ rad (ion), and the best six-gate design is $1.9\%$ worse
rather than better at $19.2$ rad. The obstruction is the condition number of the
Jacobian, not the gate count: saving gates buys a factor of a few, not the two to
three orders of magnitude that are missing.

At $n=7$ the absolute figure cannot be quoted at all. The row norms span $2.1
\times10^{21}$ while the row-balanced condition number of the quoted five-gate
design is $1.9\times10^{4}$; the absolute one is
$\kappa_{\mathrm{abs}}=2.2\times10^{11}$ for the quoted design ($9.3\times10^{11}$
for the five-gate design of the sweep), from $\sigma_{\max}=7.55$ and a
$\sigma_{\min}$ resolved to a few digits only, so the number is a statement about
this computation rather than a certified bound. The floor it implies is not
$\kappa_{\mathrm{abs}}\sigma_{\mathrm{res}}$ but the rms of the marginals,
$\sigma_{\mathrm{res}}\bigl[(1/m)\sum_{i}\sigma_{i}^{-2}\bigr]^{1/2}$ with
$m=4n-1=27$, which for $\sigma_{\mathrm{res}}=10^{-3}$ is $6.7\times10^{6}$\,rad
for the quoted design: all $27$ parameters of a seven-qubit repetition code
cannot be recovered by this protocol at any shot count. What we can report is the row-normalized metric,
where the rms falls monotonically from $8.08$ at five gates to $7.52$ at seven
and has not saturated. Unlike $n=3$, the $n=7$ scan has not reached its turning
point, so we make no claim about where it is.

\subsection{A second axis: the reference channel at every site}
\label{sec:level0}

Every count above assumes the reference channel is measured on site~$1$ only, as
in the companion's fingerprint, whose single level-0 block is the $i=1$ member of
the set used here. Measuring it at \emph{every} site gives a
$16(n+2)$-component map $L_{n}$---$L_{3}=80$, $L_{5}=112$, $L_{7}=144$---and needs
no recovery gate, because no syndrome is resolved: the two code blocks enter as
logical channels rather than as branches, so there is no $W_s$ to apply. What it
keeps from the code is the deficit removal below, without which it would stop at
rank $3n$. The $112$ here is $16(5+2)$, not the
$112=7\times16$ of Table~\ref{tab:floor}, which counts four second-order code
blocks; likewise $L_{7}=144$ is a component count, whereas the $144$ settings of
the companion's two-qubit marginal are $16$ inputs times $9$ settings. The structure is a counting argument with a proof in two lines. A
reference channel is a single-qubit unitary channel with three real parameters,
and a coupling reaches it only as a diagonal phase: with the other qubits
prepared in $|0\cdots0\rangle$ and traced out, the couplings multiply the site's
reduced Kraus operators by scalars of unit modulus, so the channel is
conjugation by
\begin{equation}
V_i=U_i(b_i,c_i,d_i)\,e^{-is_iZ},\qquad
s_i=\arctan t_{i-1}+\arctan t_i ,
\label{eq:gauge}
\end{equation}
with $t_0=t_n=0$---a single-qubit unitary. Two things follow. Because every
$V_i$ is unitary, the derivative columns of block $i$---four at an end block
and five in the interior, one per free angle---all lie in the three-dimensional
tangent space of single-qubit unitary channels, so $n$
blocks span at most $3n$, while the family has $4n-1$ parameters, so the deficit
is at least $n-1$. And the deficit is attained: multiplying $U_i$ and $U_{i+1}$
on the right by $e^{-i\tau Z}$ and decreasing $\arctan t_i$ by the same $\tau$
changes $s_i$ and $s_{i+1}$ by $-\tau$ and leaves $V_i$, $V_{i+1}$ and every
other site's channel unchanged, so each of the $n-1$ bonds carries one direction,
supported on the six angles of its two sites and on the coupling between them.
Those directions are independent, having disjoint coupling coordinates. At the
certified point below the $16n$-row level-0 block has rank $3n$---$9$, $15$,
$21$---and adding the bit and phase channels gives $11$, $19$, $27$: the code
blocks do not trace out the neighbours, so they see the gauge and supply exactly
the deficit. Hence $B^{*}=0$ at every $n$ for the enlarged map at the certified point; at
$\theta=0$ the same map drops to rank $3n+1$, which is the degeneracy discussed in Sec.~\ref{sec:floor}. This does not
revise Table~\ref{tab:rank}: that table counts recovery gates for the
three-component map and remains correct for it, while the statement here is
conditional on measuring $n$ channels the three-component protocol does not
measure, each a $16$-setting single-qubit tomography with no syndrome branch.

Rank is not precision, and measuring the reference everywhere does not make the
gates redundant: wherever both axes are opened the joint point is the best of the
three, and what changes with $n$ is only whether it reaches the target. At the
minimum-gate design---$\varnothing$, one and five gates at $n=3$, $5$, $7$---the
three-component floor is $6.29$\,rad at $n=3$, $3.4\times10^{3}$\,rad at $n=5$
and $6.7\times10^{6}$\,rad at $n=7$, all shot-independent. The $n=5$ entry is a no-discard value: the two
smallest Fisher eigenvalues of that design sit at the round-off floor of the
computation, and a threshold that drops them would report $37$\,rad instead, so
we quote the unthresholded number and the rank statement below rather than the
thresholded one. The rank of $24$ of $27$ is a property
of full decoding, a different design, and there three directions are unreachable
at any shot count. With the reference on every site and no gate at all the same
three points give $5.2\times10^{-2}$, $0.17$ and $6.8$\,rad on sc-cur: a factor
$121$ at $n=3$, $2\times10^{4}$ at $n=5$, and at $n=7$ a finite but useless number, the
obstruction there changing in kind rather than in size---$\kappa_{\mathrm{abs}}$
falls to $2.6\times10^{5}$, and the floor stays three orders above target. Opened
together the two axes give $2.55\times10^{-3}$\,rad at $n=3$---$6.5$ times below
the same three-gate design on three components and $20$ times below the gate-free
enlargement, and below target---and $6.9\times10^{-2}$\,rad at $n=5$ (no bit
correction, phase on sites $1$ and $2$), a further $2.5$ times below the gate-free
value but $7$ times above target. At $n=7$ neither axis is within reach and we
report no joint point. The mirror collision is untouched at $n=3$ (residual
$2.2\times10^{-16}$ in the $80$-component $|0\cdots0\rangle$-frame map), so this enlargement leaves the reflection ambiguity in place
and it must be removed separately: the two-qubit marginal of
Sec.~\ref{sec:mirror} does so, and so does the companion's single-qubit
$|+\rangle$-reference block, at twelve settings.

What the extra channels cost is precision, not gates. Counted in the same unit, the gate-optimal
design needs $16(1+2^{n})=144$, $528$ and $2064$ settings at $n=3$, $5$, $7$,
against $16(n+2)=80$, $112$ and $144$ here---$44\%$, $79\%$ and $93\%$ fewer, the
saving growing with $n$ because the $2^{n}$ branch structure is exactly what the
$n$ reference blocks replace. A variant that keeps all $2^{n}$ branches and also
measures the reference at every site---the extended fingerprint of
Sec.~\ref{sec:cost}---costs $16(n+2^{n})$: $+22\%$, $+12\%$ and $+4.7\%$ over
$16(1+2^{n})$, and it is a different design from the one counted here. The
$16$ is the setting count per block used throughout this paper; the companion
builds its references from twelve settings, four inputs times three bases, and in
that convention the three savings are $51\%$, $82\%$ and $94\%$, so the
sixteen-setting count is the conservative one and no conclusion below depends on
which is used. In the
cost model of Sec.~\ref{sec:cost}, which counts recovery gates, the extra
channels cost none, and in the setting count they cost less. What they cost is
precision: at $n=3$ a three-gate design still beats their floor by $3.1$ times at
$1.8$ times the settings, and at $n=5$ the two-axis point by $2.5$. What they do
not buy is correction: the enlarged map is a characterisation map, not a
corrected one.

The ranks are exact over $\mathbb{Q}$, not thresholded. At a rational point of
the same model---$a_i$ and $s_k=\sqrt{1+t_k^2}$ rational, denominators at most
$144$, not the point of Sec.~\ref{sec:rank}---the $16(n+2)\times(4n-1)$ Jacobian
is assembled from analytic derivatives
and eliminated over $\mathbb{Q}$ with no tolerance: rank $3n$ with $n-1$ integer
kernel vectors of support seven, a nonzero integer minor of order $3n$ ($249$,
$788$ and $1595$ bits at $n=3,5,7$), then rank $4n-1$ with the two code channels
and a nonzero minor of that order ($668$, $2171$ and $4703$ bits), and the same
ranks at $p=65521$ and $p=65519$ (\texttt{rank\_exact.py} in the archive).

At the paper's certified point the level-0 Jacobian of the same map has
$\sigma_{3n}/\sigma_1=0.46$--$0.50$ against
$\sigma_{3n+1}/\sigma_1=2.4\times10^{-16}$, the level of machine precision, so
rank $3n$ is a gap there and not a tolerance; the $2.5\times10^{-11}$ that a central difference at
$\epsilon=10^{-5}$ reports in that direction is the stencil's truncation floor,
falling as $\epsilon^2$---$2.3\times10^{-5}$, $2.3\times10^{-7}$ and
$2.3\times10^{-9}$ at $\epsilon=10^{-2}$, $10^{-3}$, $10^{-4}$---and not a
property of the map. The certified point is written out once, at $n=3$, for the reader who wants to check it:
$\theta=(\tfrac15,\tfrac25,\tfrac25,\tfrac25,\tfrac15,\tfrac25,\tfrac25,\tfrac25,\tfrac15,\tfrac{9}{40},\tfrac{20}{99})$
in the site-major order
$(\alpha_1,\beta_1,\delta_1,\alpha_2,\beta_2,\delta_2,\alpha_3,\beta_3,\delta_3,\gamma_{12},\gamma_{23})$
for this point; the same rational construction gives the certified point at each
$n$. The companion lists $\theta$ in block
order, $\alpha_1\alpha_2\alpha_3/\beta_1\beta_2\beta_3/\delta_1\delta_2\delta_3$; the two readings coincide
at $n=3$ but not at $n=5$ and $7$, so the order matters when the certificate is
checked against the companion's code. Its entries exceed $0.1$~rad, so it lies outside
the sampling box $[-0.1,0.1]^{9}\times[-0.05,0.05]^{2}$ used for the statistics above: the
certificates are independent of that box rather than supported by it. Nothing in the
argument is specific to the repetition code:
it needs only that each site's reference channel carry three parameters
independent of the couplings, which is what makes the deficit $n-1$; the repair
uses the fact that this family has exactly two code blocks.

\section{A hard floor for all-parameter recoverability}
\label{sec:floor}

Section~\ref{sec:precision} compares designs at a fixed point, and its most portable
statement is an ordering: three gates beats six at every point sampled. An ordering
is not a guarantee. The factor attached to it belongs to the point---the condition
number spans a factor $600$ over the sampled set alone---and the placement that wins
must be chosen once $\theta$ is known, which is exactly what an experimenter who has
to build a decoder cannot do. We therefore ask the complementary question. If the
design is fixed in advance, what holds at \emph{every} point?

The answer is a floor rather than an optimum, and its price is fixed before the
experiment starts. Run two designs, $a$ and $b$, splitting a total budget $N$
evenly. Here $a$ and $b$ are arbitrary members of the $64$-design family of
Sec.~\ref{sec:model}; $A$, $B$ and full decoding denote the three designs of
Sec.~\ref{sec:precision} throughout. The two data sets are independent, so their
Fisher informations add \cite{kay},
\begin{equation}
\mathcal{F}_{\rm stack}=\frac{N}{2}\bigl(J_a^{T}J_a+J_b^{T}J_b\bigr),\qquad
\mathrm{Cov}_{\rm stack}=\frac{2}{N}\bigl(J_a^{T}J_a+J_b^{T}J_b\bigr)^{-1}.
\label{eq:stack}
\end{equation}
Equation~\eqref{eq:stack} uses a single symbol $J_s$ for the weighted Jacobian
$\eta_s J_s/\sigma$ of Sec.~\ref{sec:precision}, so that a design at $N_s$
repetitions carries information $N_s J_s^{T}J_s$ exactly. The visibility $\eta$
depends only on the gate count, and the component error $\sigma$ carries no
component index, so for two three-gate designs $J_s$ differs from the bare
Jacobian by one common factor; that factor cancels from every ratio below, and a
positive factor cannot change a rank.
Two consequences follow from Eq.~\eqref{eq:stack} alone: no noise table, no working
point, no sampling.

\begin{proposition}[Union, not intersection]
\label{prop:union}
For any designs $a,b$ and any $\theta\in\Th$,
$\mathcal{F}_{\rm stack}-\tfrac{N}{2}J_a^{T}J_a=\tfrac{N}{2}J_b^{T}J_b\succeq0$, and
likewise with $a$ and $b$ exchanged. Hence the pair is locally identifiable wherever
either member is, and its Jacobian $[J_a;J_b]$ has at least the rank of the better
member.
\end{proposition}

\begin{proof}
$J_b^{T}J_b\succeq0$; adding it to $\tfrac N2J_a^{T}J_a$ preserves positive
definiteness.
\end{proof}

\begin{proposition}[The premium is bounded]
\label{prop:premium}
For designs whose Jacobians have full rank,
$\mathrm{Cov}_{\rm stack}\preceq2\,\mathrm{Cov}_a$ and
$\mathrm{Cov}_{\rm stack}\preceq2\,\mathrm{Cov}_b$ in the Loewner order
\cite{horn}. Consequently, for the rms over the $4n-1$ parameters and for the
largest marginal standard deviation alike,
\begin{equation}
\mathrm{rms}_{\rm stack}\le\sqrt2\,\min_{s=a,b}\mathrm{rms}_s .
\label{eq:premium}
\end{equation}
\end{proposition}

\begin{proof}
$\mathrm{Cov}_{\rm stack}=2(J_a^{T}J_a+J_b^{T}J_b)^{-1}\preceq
2(J_a^{T}J_a)^{-1}=2\,\mathrm{Cov}_a$, because inversion reverses the Loewner order
on positive definite matrices; the same argument gives the bound with $b$. Both
functionals are monotone in that order, so the matrix inequality passes to them.
\end{proof}

\begin{corollary}[Cost of the insurance]
Matching the precision of the better member at budget $N$ costs at most $2N$ shots,
since Eq.~\eqref{eq:premium} gives $\mathrm{rms}_{\rm stack}(2N)\le
\min_s\mathrm{rms}_s(N)$. A factor $\sqrt2$ in deviation, and at most a factor two
in shots, is paid in advance; the gain is not, and is where the value is.
\end{corollary}

Three qualifications belong with the statement. The bound is tight: at the
structured point with only the two $ZZ$ couplings present,
$\mathrm{sd}_{A+B}/\min(\mathrm{sd}_A,\mathrm{sd}_B)=1.4142$, the ceiling to five
figures, because the parameter that limits $B$ is one to which $A$ contributes
nothing, so the split costs $B$ the full $\sqrt2$. The
right-hand side of Eq.~\eqref{eq:premium} is a minimum over members evaluated at the
\emph{same} $\theta$, so the pair inherits the better design point by point without
being told which one it is; a single fixed design has no analogue of that guarantee,
and one of the two is $719$ times worse than the other at a point found below. And
the statement concerns variance only, as the next paragraph makes exact.

\paragraph{What it does not buy.}
If the residual calibration error lies in the same $4n-1$ family---a residual
under-rotation \emph{is} an error in $(\alpha,\beta,\delta)$---then every design is
displaced by $J_s\,\delta\theta_{\rm cal}$, and the generalized least-squares
estimate returns $\delta\theta_{\rm cal}$ for the stack as well. The bias is
therefore independent of the number of designs while the deviation falls as
$\sqrt2$: its $z$-score grows. The $10$\,mrad statement of Sec.~\ref{sec:cost}
remains conditional on the calibration budget, not on the gate budget. Neither does
the floor repair the degeneracies of the model. On the stratum with
$\alpha=\beta=\gamma=0$ no design of the $64$ exceeds rank $10$, and at $\theta=0$
all reach rank $5$; Proposition~\ref{prop:union} cannot help there, because there is
no direction for any member to contribute.

\paragraph{Numbers.}
The Loewner inequality of Eq.~\eqref{eq:premium} was evaluated rather than assumed.
Three fingerprints are reported. The $48$-component one is Eq.~\eqref{eq:fp}; the
$80$-component one adds the reference channel of Sec.~\ref{sec:level0} on sites
$2$ and $3$; the $112$-component one appends four further code blocks---the
second-order channels $R_{XX}$, $R_{XZ}$, $R_{ZX}$ and $R_{ZZ}$ that correct a
weight-two syndrome, $16$ components each---so that $112=7\times16$ against
$48=3\times16$. That was the construction of the earlier versions of
\cite{companion}; the present version replaces the four second-order code blocks
by two $|+\rangle$-frame references, so its $112$ is a different seven-block map
and only the $48$- and $80$-component columns below are common to the two papers.
The last is included
because a parallel design can measure all seven blocks in one pass, and the point
of the table is that the floor statement depends on none of them.
At $201$ points in the $48$-component fingerprint, $61$ in the $80$-component one
and $41$ in the $112$-component one---the working point plus an independent
random draw, separate from the sweep below---the largest eigenvalue of
$\mathrm{Cov}_{\rm stack}-2\,\mathrm{Cov}_a$ scaled
by $|\mathrm{Cov}_a|_{\max}$ is $-3.7\times10^{-8}$, $-2.2\times10^{-4}$ and
$-9.0\times10^{-9}$, its
trace excess is negative ($-1.8\times10^{2}$, $-3.1$ and $-3.0\times10^{1}$), and none of
the $303$ points loses rank. Table~\ref{tab:floor} reports the same quantities over
a larger draw of $300$ uniform working points in $\Th$---uniform in the parameter
box itself, $[-0.1,0.1]^{9}\times[-0.05,0.05]^{2}$ at $n=3$, which is the measure
that makes ``uniform in $\Th$'' concrete and is the one the reproduction package
uses (seed $20261115$).

\begin{table}[t]
\centering
\small
\caption{The floor over $300$ uniform working points in $\Th$: two designs, total
budget $N$ split evenly, every entry a ratio so that $N$ cancels. $r$ is the ratio
of Eq.~\eqref{eq:premium} and cannot exceed $\sqrt2=1.4142$; ``same shots'' compares
the pair at budget $N$ with the better single design at $N$, ``same gates'' with
design $B$ at $2N$, so that both use the same number of single-qubit Pauli gates and
the ceiling is $2$. The $80$-component column is $L_{3}=16(3+2)$ of
Sec.~\ref{sec:level0}; the $112$-component one is $7\times16$---three basic and
four second-order code blocks---and is neither $L_{5}$ nor the seven-block
fingerprint of the present version of \cite{companion}. The rms criterion is
the one of Sec.~\ref{sec:precision}; the
last row repeats the leading one in the stricter max-marginal metric.}
\label{tab:floor}
\footnotesize
\setlength{\tabcolsep}{4pt}
\begin{tabular}{lccccccccc}
\toprule
& \multicolumn{3}{c}{$48$ components} & \multicolumn{3}{c}{$80$ components}
& \multicolumn{3}{c}{$112$ components}\\
\cmidrule(lr){2-4}\cmidrule(lr){5-7}\cmidrule(lr){8-10}
& median & 95th & worst & median & 95th & worst & median & 95th & worst\\
\midrule
$r$, same shots & $0.658$ & $0.896$ & $1.178$ & $1.084$ & $1.274$ & $1.334$
& $0.583$ & $0.846$ & $0.997$\\
share with $r<1$ & \multicolumn{3}{c}{$99.3\%$}
& \multicolumn{3}{c}{$13.3\%$} & \multicolumn{3}{c}{$100\%$}\\
$r$, same gates & $0.725$ & $1.159$ & $1.666$ & $1.370$ & $1.728$ & $1.859$
& $0.591$ & $1.061$ & $1.346$\\
\midrule
$r$, same shots, max marginal & $0.584$ & $0.987$ & $1.2547$ & $1.088$ & $1.286$
& $1.353$ & $0.531$ & $0.971$ & $1.2280$\\
\bottomrule
\end{tabular}
\end{table}

In the rms metric the pair is below the better single design at $99.3\%$ of the
points ($48$ components) and at every point in the $112$-component fingerprint, with
median ratio $0.658$ and $0.583$: median gains of $1.52$ and $1.71$ in deviation,
$2.3$ and $2.9$ in shots. The ceiling is respected with room---the worst ratio is
$1.178$ and $0.997$ against $\sqrt2$---and the stricter max-marginal metric
($\maxsd$, the largest marginal standard deviation, which is what the supplementary
note tabulates), behaves the same way with a longer tail
($96.7\%$ and $96.3\%$ of points; worst $1.2547$ and $1.2280$). Compared at equal
\emph{gate} count, the pair at $N$ against design $B$ at $2N$, the median gain falls
to $1.38$ and $1.69$ and the pair loses at $16\%$ and $8\%$ of the points, by at
most $1.67$ and $1.35$---below the ceiling $2$ that the counting argument allows.
The $80$-component map is of a different kind: there the pair is \emph{worse}
than the better single design at $87\%$ of the points, with median ratio $1.084$
(a net loss of $8\%$) and worst $1.334$, and at equal gate count it is worse at
$88\%$ of them (median $1.370$, worst $1.859$).

\paragraph{Why the premium shrinks.} Stacking buys nothing on a block that does
not depend on the recovery design, and the two members measure the reference
channels identically. The median ratio orders by that fixed share of the
fingerprint: one block of seven fixed gives $0.583$, one of three gives $0.658$,
and three of five gives $1.084$. The guarantee of Proposition~\ref{prop:union} is
untouched---the pair still loses no direction and still costs at most $\sqrt2$---but in
the $80$-component map the channels that carry the identification are the ones
both designs already share, so there is little left for a second design to
insure. Buying reference channels on sites $2$ and $3$ and buying a second
placement are, to first order, substitutes: Sec.~\ref{sec:level0} prices the
first, this section the second.

The equal-shot row carries a visibility asymmetry that a unit per-component error
hides and that one should not mistake for an ignored gate error: the pair applies
six Paulis where the better member applies three, so its $\eta$ is lower by
$(1-2p_1)^{3}$, which is $0.6\%$ on sc-cur and $0.2\%$ on sc-best---a correction
to the variance of order one percent, two orders below the factor $1.5$ the row
reports. The gate error itself is in the model, through Eq.~\eqref{eq:eta}.

Random draws do not find the bad points, so we searched: $3000$ draws followed by
Nelder--Mead refinement of $\log\maxsd$ from the four best starts. Two of the three
single designs are \emph{rank deficient} at their own worst point---design $B$ and
full decoding do not merely lose precision there, they lose a parameter---while the
pair stays at $12.6$ and $8.73$ ($48$-component fingerprint), below the better member there
by factors $2.9$ and $9.5$.
At the worst point of design $A$ the asymmetry is starker: $A$ stands at
$6.98\times10^{4}$, $B$ at $97.1$ and the pair at $37.1$---better than $A$ by
$1880$ and than $B$ by $2.6$, against a ceiling of $137$ at that point.

Two further checks delimit the claim. On the $31$ support strata obtained by zeroing
every nonempty subset of the five parameter groups, the pair attains the best
single-design rank on all $31$, and on one of them---$\alpha=\beta=\delta=
\gamma_{23}=0$---exceeds it, rank $9$ against $8$; on the $8$ strata where the model
itself is degenerate no design does better, which is the boundary. And a third
design is not worth its gates: at the working point of the companion paper
\cite{companion} the second design improves the deviation by factors $1.60$
($48$ components) and $2.16$ ($112$), while a third adds only $1.11$ and $1.04$. Two designs also sit at the right price: the pair costs six gates, the
same as full decoding and twice the three-gate budget that
Sec.~\ref{sec:precision} identifies as the optimum.

The floor changes what can be promised rather than what can be achieved. At that
working point, at equal total shot budget and one repeat per branch, the pair's rms
is $3.64$ against $6.07$ for design $B$, $6.63$ for $A$ and $13.61$ for full
decoding, with a unit per-component error---so the same deviation is reached with
$2.8$, $3.3$ and $14$ times fewer shots---and it does so with nothing to choose in
advance.

\paragraph{Reaching a target.}
A precision surplus is worth nothing unless it converts into shots, and the
conversion is where a badly conditioned point does its damage, since
$\maxsd\propto N^{-1/2}$: a target of $0.01$\,rad becomes a shot count by
squaring. At a unit per-component error the pair reaches it in
$7.5\times10^{4}$ shots at the median of the $300$ draws
($5.5\times10^{5}$ in the $48$-component fingerprint), in $4.6\times10^{5}$ and
$2.9\times10^{6}$ at $95\%$ of them, and at worst in $2.5\times10^{6}$ and
$1.4\times10^{7}$. At those two worst points the better single design, chosen
there with $\theta$ known, needs $4.9$ and $6.9$ times more, and full decoding
$70$ and $10$ times more. The maxima are sample statistics, and they have to be:
no finite maximum over $\Th$ exists. The box contains $\theta=0$, where every
design of the $64$ has rank $5$, and the neighbourhoods of the degenerate strata
counted above, so $\maxsd$ diverges inside $\Th$ and every finite budget
describes a region rather than the domain. Nor are the sample maxima tight for
their own region: a greedy coordinate ascent from the worst drawn point,
confined to $\Th$, multiplies the count by $4.7\times10^{3}$ ($112$ components)
and $1.1\times10^{4}$ ($48$), with all three designs at rank $11$---and at the
$48$-component point so reached, design $B$ alone is cheaper than the pair by
$1.6$, while at the $112$-component one full decoding is cheaper by $2.8$.
Through all of it the pair stays inside the factor of Eq.~\eqref{eq:premium} of
its own members, and that inequality is the one statement that holds at every
point of $\Th$; what a working point costs is a property of the region it sits
in. The counts assume a unit per-component error; a platform's per-shot error
rescales them by its inverse square, and Sec.~\ref{sec:cost} carries that
factor. Proofs, figure and recorded numbers are in the supplementary note
\cite{floor}.

\section{Cost of an actual experiment}
\label{sec:cost}

\begin{table}[t]
\centering
\caption{Three-qubit recipe with the three-gate design
($S_{\mathrm{bit}}=\{3\}$, $S_{\mathrm{ph}}=\{1,2\}$) at the reference working
point of Table~\ref{tab:n3}. ``Floor'' is the shot-independent limit set by
$\sigma_{\mathrm{res}}$; the next column rescales the budget to a fixed $10^{10}$
total shots. Budgets are Eq.~\eqref{eq:ntot}: $144$ configured shots per unit of
$N$. Wall clock is $N_{\mathrm{tot}}$ times the per-shot time
$t_{\mathrm{cycle}}+8\times60\,\mathrm{ns}+3\times25\,\mathrm{ns}$, which is
$5.555\,\mu$s on the superconducting tables and $200.555\,\mu$s on the ion
table. $p_2$ is not included; restoring it in Eq.~\eqref{eq:eta} raises the two
reachable budgets by $19\%$ (sc-best) and $3\%$ (ion), i.e.\ to
$1.51\times10^{8}$ and $1.01\times10^{8}$ shots.}
\label{tab:cost}
\begin{tabular}{lcccc}
\toprule
platform & floor (rad) & rms at $10^{10}$ shots & shots for $10$\,mrad & wall clock\\
\midrule
sc-cur & $1.66\times10^{-2}$ & $1.66\times10^{-2}$ & unreachable & ---\\
sc-best & $4.91\times10^{-3}$ & $5.01\times10^{-3}$ & $1.27\times10^{8}$ & $707$\,s\\
ion & $1.62\times10^{-3}$ & $1.89\times10^{-3}$ & $9.75\times10^{7}$ & $1.96\times10^{4}$\,s\\
\bottomrule
\end{tabular}
\end{table}

\noindent\textit{Extended fingerprint.} The level-0 channels cost $176N$
shots at $n=3$ ($32N$ more than Eq.~\eqref{eq:ntot}, no extra gates). The floor is
$2.55\times10^{-3}$ rad: $10$ mrad takes $3.21\times10^{6}$ shots and $0.1$ rad
$3.00\times10^{4}$ (from $1.02\times10^{6}$). At equal \emph{total} shots $T$
($T/144$ over $48$, $T/176$ over $80$), the rms gain is $5.7$--$6.4$---not the
smaller per-channel companion number. Rank and mirror collision are unchanged at
this $n$; Sec.~\ref{sec:level0} gives the structure at all three.

The three-gate design is $4$ checks (two $Z_iZ_{i+1}$, two $X_iX_{i+1}$), $8$
controlled-X gates, $2$ ancillas, $16$ settings, $9$ branches (four bit, four
phase, one reference) and three Paulis per Eq.~\eqref{eq:Ws};
Table~\ref{tab:cost} gives what it buys.

\paragraph{Counting shots.}
$N$ in Eq.~\eqref{eq:sigma} is shots \emph{per component, per setting, per
branch}; with $16$ settings, one reference branch and $2^{n-1}=4$ syndrome
branches per code,
\begin{equation}
N_{\mathrm{tot}}=16\,(1+2^{n})\,N=144\,N
\label{eq:ntot}
\end{equation}
configured shots; a code-block component collects $N/4$ per sector after a
split, as in Eq.~\eqref{eq:stack}. The rms multiplier
$\sqrt{A}$, with $\mathrm{rms}^{2}=A\sigma^{2}+\mathrm{floor}^{2}$, is $A=67.1$
($\sqrt{A}=8.19$) at the reference point and $A=65.9$ on the ion table, where the
visibility differs. The naive amplification
$\mathrm{floor}/\sigma_{\mathrm{res}}=16.4$ assumes shot noise on all $48$
components equally; a code-block component carries a quarter of the
reference-block shots, so the correct multiplier is $8.19$---a factor $4$ in the
budget. The arithmetic is one line:
\begin{align}
\sigma^{2}&=\frac{10^{-4}-\mathrm{floor}^{2}}{A}
=\frac{10^{-4}-2.41\times10^{-5}}{67.1}=1.13\times10^{-6},\notag\\
N&=1/\sigma^{2}=8.8\times10^{5},\qquad
N_{\mathrm{tot}}=144\,N=1.27\times10^{8}.
\label{eq:oneline}
\end{align}
A shot carries its branch's circuit:
$t_{\mathrm{shot}}=t_{\mathrm{cycle}}+8\times60\,\mathrm{ns}+g_P\times25\,\mathrm{ns}$
for the eight checks and $g_P$ Paulis (nominal $60$\,ns and $25$\,ns), i.e.\
$5.555\,\mu$s for three gates ($5.630$ for six) and $200.555\,\mu$s on the ion
table. Then $1.27\times10^{8}\times5.555\,\mu\mathrm{s}=707$\,s and
$9.75\times10^{7}\times200.555\,\mu\mathrm{s}=5.4$\,h. The $10\%$ over the bare
$5\,\mu$s cycle is the $0.555\,\mu$s of circuit, not an overhead.

The reading is platform dependent, working-point dependent, and none of it is a
formality. With the three-component fingerprint of Eq.~\eqref{eq:fp}, at the
reference point on a current superconducting device the target is unreachable,
and the reason is not statistics: the floor is set by
$\sigma_{\mathrm{res}}=10^{-3}$, so reaching $10$ mrad requires pushing the
calibration residual below $6.5\times10^{-4}$. That is a calibration problem, and
no number of shots substitutes for it.

The extended fingerprint changes that verdict at the same point (Table~\ref{tab:cost}
is the three-component map): the added blocks constrain $q_2,q_3$ angles the
single reference barely sees, their standard deviations falling from
$15$--$26$ to $0.3$--$4.3$\,mrad ($\sim0.3$ for most), with no new gates or rank
change.

On a best-case superconducting device the target takes $1.27\times10^{8}$ shots
in $12$ minutes; a trapped ion needs \emph{fewer} ($9.75\times10^{7}$) but $5.4$
hours, its cycle time being $40\times$ longer. The $28$-fold wall-clock gap is
bought with a factor $3$ in the floor and only $1.3$ in shots---the ion table
pays in time per shot, not statistics. With $p_2$ restored the entries become
$1.51\times10^{8}$ in $14$ minutes and $1.01\times10^{8}$ in $5.6$ hours, and the
wall-clock ratio falls to $24$.

Table~\ref{tab:points} also puts three gates on the current device: at the
companion's working point the floor is $4.9$\,mrad, below target with the
residual left where it is, and $10$ mrad costs $1.2\times10^{7}$ shots in
$65$\,s. The factor between the two points ($33$ at full decoding) exceeds the
factor between three gates and six at either point ($2.3$ to $22$): at this scale
the gate budget is not what stands between the protocol and $10$ mrad.
Conditioning is, and conditioning is a matter of where the sample sits and how
well the residuals are calibrated, not of the decoder.

\begin{figure}[t]
\centering
\includegraphics[width=\linewidth]{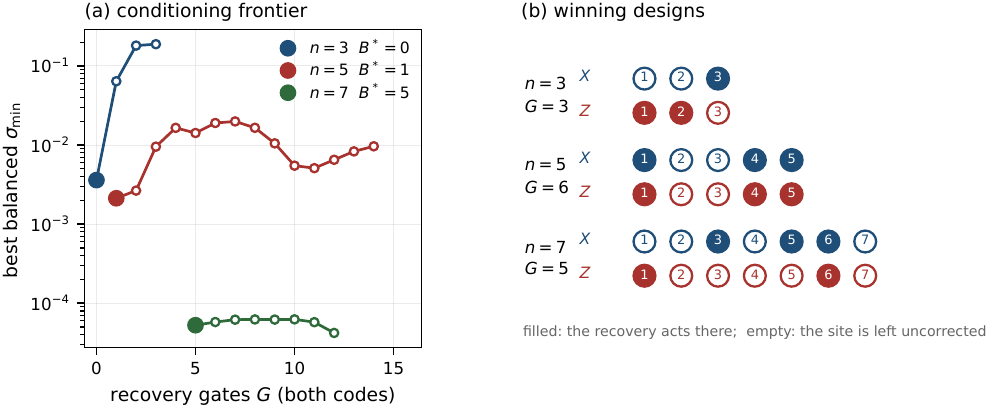}
\caption{(a) Best row-balanced smallest singular value at each gate budget. The
filled markers are the minimal full-rank counts $B^{*}$. At $n=3$ the curve rises
from $B^{*}=0$ and saturates by three gates; at $n=7$ it is flat and low well
beyond $B^{*}$, which is why the five-gate design is full rank but poorly
conditioned in the absolute metric. (b) The quoted designs. Filled sites carry
the recovery Pauli; empty sites are left uncorrected. The two codes use
different site sets at $n=3$ and $n=7$, and that decoupling is where the saving
comes from.}
\label{fig:design}
\end{figure}

\section{The designs are not covariant under the chain reversal}
\label{sec:mirror}

A repetition code on a chain has a reflection symmetry, and the companion paper
\cite{companion} classifies the fingerprint ambiguities it generates: the code
channels are exactly invariant under the site reversal $R:i\mapsto n+1-i$, and
the resulting degeneracy is removed there either by appending one two-qubit
marginal or, at twelve settings, by changing the preparation of the two
spectator qubits, at a stated cost. That construction assumes the recovery is itself covariant. Our
optimal designs are not, and the interaction deserves an explicit statement
rather than a citation.

\begin{lemma}[Covariance is what transfers the invariance]
\label{lem:cov}
Let $R$ act on sites, and let $W$ be a recovery. If
$W_{Rs}=RW_sR^{\dagger}$ for every syndrome $s$, then the code blocks satisfy
$\R_{c}(R\theta)=\R_{c}(\theta)$ for all $\theta$. If the condition fails, the
code blocks are not invariant in general.
\end{lemma}

\begin{proof}
$K_{R\theta}=RK_\theta R^{\dagger}$ because the generator is reflection
covariant, and the code space, the syndrome projectors and the logical algebra
are carried onto themselves by $R$. Under the hypothesis, the Kraus operators
$V^{\dagger}W_sP_sK_\theta V$ map onto $V^{\dagger}W_{Rs}P_{Rs}K_{R\theta}V$, so
the channels coincide. Without the hypothesis the two Kraus sets differ
generically.
\end{proof}

The test is local: $RW_sR^{\dagger}=W_{Rs}$ has residual $0$ for the empty and
full site sets at every $n$, and residual $2$ (the norm of $Z_1-Z_3$) for any
partial set not invariant under $R$---exactly our optima. At $n=3$ the winning
design corrects $\{3\}$ (bit) and $\{1,2\}$ (phase), neither reflection invariant.

\begin{table}[t]
\centering
\caption{The companion paper's compensated collision point, evaluated with
different decoders. $\theta$ and $S\theta$ are the two in-domain points whose
$|0\cdots0\rangle$-frame fingerprints---three references and two code channels,
$80$ components---coincide under the standard symmetric recovery.
The last column is the Jacobian rank at $\theta$. Every non-covariant allocation
tested gives $1.33\times10^{-3}$ in magnitude; the sign depends on the
allocation.}
\label{tab:mirror}
\begin{tabular}{lccc}
\toprule
decoder ($S_{\mathrm{bit}}$ / $S_{\mathrm{ph}}$) & gates
& $\max|\Msf(\theta)-\Msf(S\theta)|$ & rank\\
\midrule
$\{1,2,3\}$ / $\{1,2,3\}$ (standard) & $6$ & $3.3\times10^{-16}$ & $11/11$\\
$\varnothing$ / $\varnothing$ & $0$ & $4.4\times10^{-16}$ & $11/11$\\
$\{3\}$ / $\{1,2\}$ (quoted design $A$) & $3$ & $1.33\times10^{-3}$ & $11/11$\\
$\{1\}$ / $\{2,3\}$ & $3$ & $1.33\times10^{-3}$ & $11/11$\\
$\{2\}$ / $\{1\}$ & $2$ & $1.33\times10^{-3}$ & $11/11$\\
\bottomrule
\end{tabular}
\end{table}

Evaluated at the collision point of \cite{companion}, with that $80$-component
map reproduced to $3.3\times10^{-16}$ for the standard decoder as a check on the
implementation, the non-covariant designs separate the mirror pair
(Table~\ref{tab:mirror}). The separation is real but modest: $1.33\times10^{-3}$
against $5.59\times10^{-2}$, which is what the companion achieves both with its
two-qubit marginal and, at twelve settings, with its $|+\rangle$ reference, a
factor $42$ weaker. It is also placement independent---the second
three-gate design of Sec.~\ref{sec:point}, bit $\{1,2\}$ and phase $\{3\}$, gives
the same magnitude, in the phase block only, at the same rank---so the verdict
below does not depend on which of the two optima an experimenter builds. Over
$20$ random mirror pairs the same picture holds
at the level of the code blocks---machine precision for the standard decoder
($4.4\times10^{-16}$) and a median of $9.8\times10^{-4}$ (up to
$\lambda=0.02$) and $2.2\times10^{-3}$ (up to $\lambda=0.05$) for the three-gate
design---while the Jacobian stays full rank at $11/11$ in every case. The
non-covariant decoder therefore breaks the degeneracy weakly and costs nothing in
identifiability.

Three consequences, in order of how much they matter.

(i) \emph{Nothing is transferred in either direction.} The companion's
identifiability statements are made for the standard decoder, where the
hypothesis of Lemma~\ref{lem:cov} holds and the theorem is correct as stated; the
certificates of Sec.~\ref{sec:rank} are made for the decoders reported here, with
the same decoder used to define the map and to test its rank. Neither set of
numbers is an evaluation of the other. In particular, the five-gate $n=7$ design
does not inherit the reflection degeneracy, and the companion's $24/27$ result at
full decoding does not apply to it.

(ii) \emph{Our result sharpens the companion's hypothesis rather than
contradicting it.} The covariance condition is not a technicality: dropping it
changes the map, and Table~\ref{tab:mirror} is a worked example of the change.
A reader who wants both results should state which decoder each belongs to.

(iii) \emph{The weak separation is not a substitute for the two-qubit marginal.}
At $1.33\times10^{-3}$ it is resolvable at the shot counts of
Sec.~\ref{sec:cost} but $42$ times closer to the noise than the marginal, and its
sign is allocation dependent, so it cannot be used as an unambiguous
disambiguator. Practitioners who adopt a gate-optimal decoder should be aware
that the mirror-invariance consistency check of \cite{companion} no longer
applies to their data.

\section{Limitations}
\label{sec:limit}

The minimum counts are exhaustive within the family of Eq.~\eqref{eq:Ws}---one
Pauli per corrected syndrome, chosen from the minimum-weight representative---and
we do not claim minimality outside it; a decoder that corrects with a
non-minimum-weight representative, or with two-qubit gates, is not covered. The
rank statements are exact but local: they certify a point in exact arithmetic,
with a second point as a recheck, and that is not a proof that the rank-deficient
locus has measure zero---it does not, since a degenerate point examined here
loses a fourth dimension, so the count of three belongs to the domain rather than
to the model. For $n=7$ the health ranking compares the $945$ full-rank
pairs at the minimal budget and does not extend to larger budgets. The $n=7$
gate scan stops at $12$, so the turning point of its precision curve is not
located, and we report monotone behaviour as an observation about the scanned
range rather than as a trend. The cost model counts gates as one single-qubit
Pauli each and does not model circuit depth, mid-circuit measurement latency,
ancilla reuse constraints beyond $n-1$, or leakage; it is a counting model, not a
compiler. The precision numbers are a Gaussian GLS estimate with a stated floor,
not a Bayesian posterior, and are more working-point dependent than the rank
statements: the sample alone spans a factor $600$ in conditioning, the two quoted
points two more orders, and no floor here bounds the protocol. What transfers is
the ordering at fixed point and the monotone floor--conditioning dependence, the
ordering itself being conditional on absolute rather than proportional noise
(Sec.~\ref{sec:precision}). Both points are extreme by construction and neither
claims the domain. The extended fingerprint improves
conditioning but does not remove the mirror collision of Sec.~\ref{sec:mirror},
and the checks behind it are sampled rather than domain-wide---$61$ points for
the Loewner comparison, $300$ for the floor. The second axis of
Sec.~\ref{sec:level0} carries the qualification that every rank statement here
carries: rank $\le3n$ and the $n-1$ kernel vectors hold at every $\theta$, while
the equality with $3n$, its repair by the two code channels and the resulting
$B^{*}=0$ are certified at one rational point per $n$, not on all of $\Th$; its
floors too are single working points, not a sweep. And the entire construction rests on
the fixed-basis premise of Sec.~\ref{sec:precision}: without it every gate count
here is zero.

\section*{Data and code availability}

The analysis is reproducible from \texttt{gate\_count\_analysis.zip}, provided
as ancillary files with this submission and deposited at Zenodo together with
the experimental supplement \cite{supp}
(\href{https://doi.org/10.5281/zenodo.23993246}{\nolinkurl{doi:10.5281/zenodo.23993246}});
a manifest inside lists a SHA-256 sum for every file. The certification of
Sec.~\ref{sec:floor}---the Loewner check, the $300$-point sweep, the worst-case
search and the support strata---travels with the supplementary note \cite{floor} in
\texttt{repro\_v1.0.zip}, whose \texttt{selftest.py} re-derives those numbers in
under a minute; its fingerprint Jacobian agrees with the Mathematica export to
$2.7\times10^{-10}$ per entry. It contains the design enumeration and the modular
rank engine;
the exact rational certificate of Sec.~\ref{sec:rank}, with the $48\times27$
integer matrix, its pivots, the nonzero $24\times24$ minor and the three kernel
vectors, so the certificate can be re-checked without re-running the
construction---\texttt{verify\_certificate.py} does that in integer arithmetic,
no floating point or numpy, in under a second; the floating-point pipeline
showing why the modular engine is needed; the noise and covariance model with
its platform table; the precision-robustness scan of
Sec.~\ref{sec:precision}; and the mirror-covariance check of
Sec.~\ref{sec:mirror}; and the second-axis results of Sec.~\ref{sec:level0} in
\texttt{level0\_80d/}---the general-$n$ engine \texttt{libn.py}, the all-site
reference block, the rank decomposition at $n=3,5,7$, the exact rank certificate
\texttt{rank\_exact.py} (rational assembly, kernel vectors, integer minor, two
primes), and \texttt{floor80.py}, which reproduces the $80$-component columns of
Table~\ref{tab:floor} on the same $300$ draws.
Every script prints its numbers, including failures, and the two fingerprint
implementations agree to machine precision on the collision point.

\section*{Disclosure}

A large language model assisted in three separable ways: it drafted the first
analysis scripts from the author's specifications (design enumeration, modular
rank engine, plotting), re-derived quoted numbers from the archived outputs as a
cross-check, and edited the language. It did not choose the problem, design
family, noise model or claims, nor produce or select any numerical result. Every
derivation, certificate and number was verified by the author, who takes full
responsibility for the content.

\end{document}